\documentclass[twocolumn,showpacs,preprintnumbers,amsmath,amssymb]{revtex4}

\usepackage{graphicx}
\usepackage{dcolumn}
\usepackage{bm}
\usepackage{amsfonts}

\begin{document}


\title{
 \textsf{Tailoring the Carrier Mobility of  Semiconductor Nanowires by
Remote Dielectrics}}

\author{Aniruddha Konar and Debdeep Jena}
\affiliation{%
Department of Physics and Department of Electrical Engineering, University of Notre Dame, Notre Dame IN 46556,USA\\
}%

\date{\today}

\begin{abstract}
The dielectric environment of thin semiconductor nanowires can affect the charge transport properties inside the wire.  In this work, it is shown that Coulomb impurity scattering inside thin nanowires can be damped strongly by coating the wire with a high-$\kappa$ dielectric.  This will lead to an increase in the mobility of free charges inside the wire. 
\end{abstract}

 \maketitle

Remarkable advances in crystal growth technology have recently enabled the fabrication of a variety of freestanding semiconductor nanostructures such as 0-dimensional (0D) nanocrystal quantum dots, 1D nanowires (NWs) and nanotubes, and 2D nanomembranes and graphene.  Charge transport properties in such nanostructures is being intensively investigated, in the hope that they might find usage in electronic and optical devices in the future.  These `bottom-up' nanostructures differ from the more extensively studied epitaxial nanostructures created by heterostructure bandgap engineering in one crucial aspect.  The dielectric environment of epitaxial nanostructures is essentially the same as the semiconducting region  (dielectric constant $\epsilon_s$) where electrons and holes reside.  However, for bottom-up nanostructures, the dielectric environment  ($\epsilon_e$) can be modified after growth. This feature offers a novel tool to engineer interactions between carriers and/or impurities by environment-mediated Coulomb interactions.  

The effect of the dielectric environment on charge transport properties of `bottom-up' nanostructures has not received much attention, as compared to its effect on optical properties \cite{keldyshJETP79}.  Recent work \cite{jenaPRL07} has shown that in 2D nanomembranes the electron mobility can be increased by 1-2 orders of magnitude by coating them with a high-$\kappa$ dielectric material.  The purpose of this work is to investigate the effect of the dielectric environment on electron transport in semiconductor nanowires. Semiconductor  NWs can now be grown with diameters of a few nanometers, which is smaller than the thermal de Broglie wavelength of the carriers, while their lengths can exceed few micrometers. At these length scales, the reduced density of states due to quantum confinement is expected to suppress scattering, and lead to high carrier mobilities \cite{sakakiJJAP80}.  Recent experiments \cite{xiangNature06} have demonstrated improved carrier mobilities in Ge/Si nanowire field-effect transistors coated with high-$\kappa$ (HfO$_{2}$) dielectrics.  

In early work on carrier transport of 1D semiconductor nanowires \cite{sakakiJJAP85,fishmanPRB86}, the effect of dielectric mismatch on transport properties was not investigated. Vagner and Mo\"{s}ko showed in their treatment of a 1D electron gas confined in a freestanding 2D membrane that the dielectric mismatch leads to a large decrease in mobility if the structure is freestanding \cite{vagnerJAP97}. In this letter, we show that for 1D nanowires, the dielectric surrounding can be used to tune the electron mobility.

\begin{figure}[t]
 \includegraphics[width=6cm]{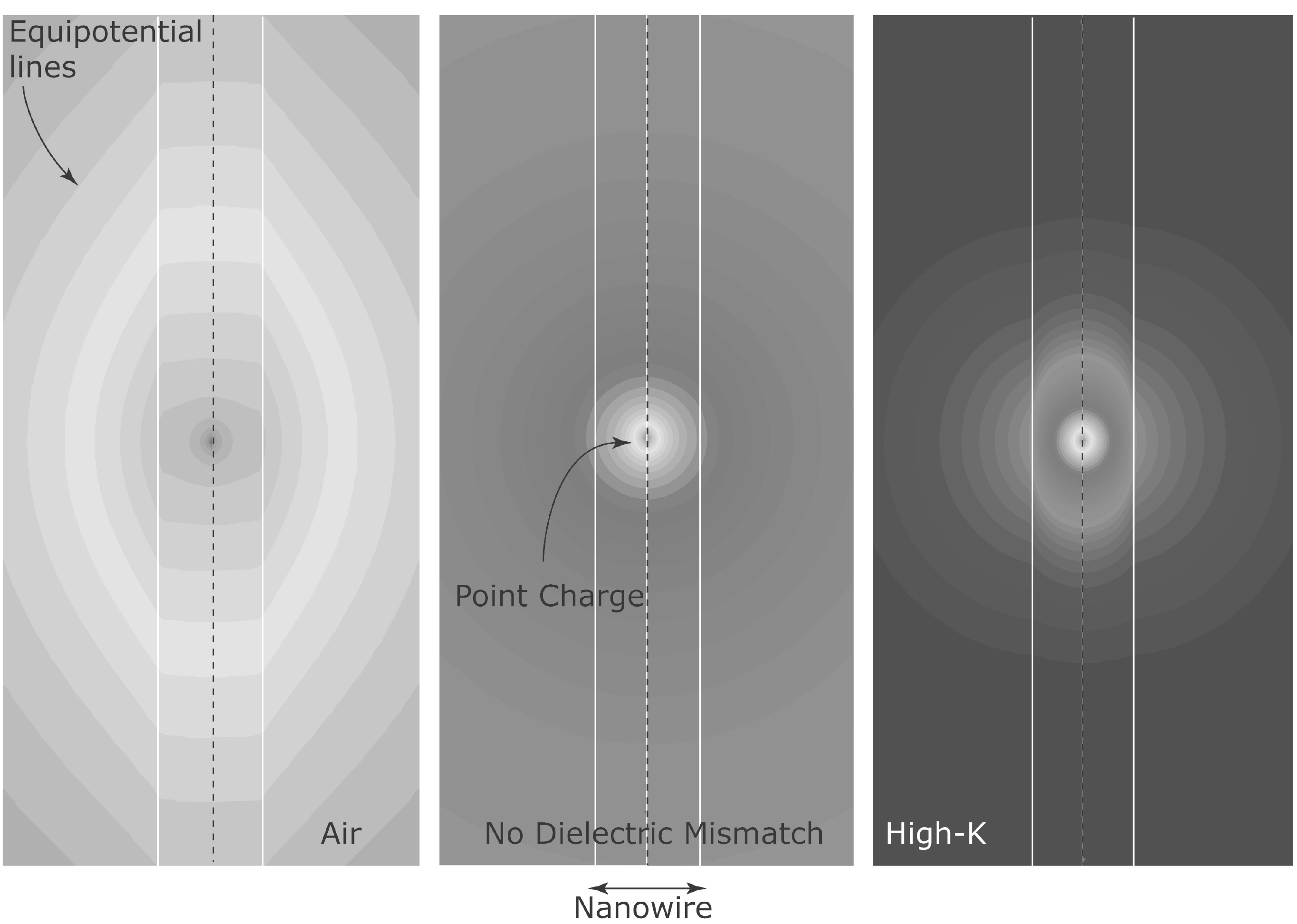}
\caption{\label{fig:epsart} Calculated Coulomb potential contours due to a point charge inside a nanowire for three different dielectric environments: $\epsilon_{e} = 1, 11, 100$, $\epsilon_{s}=11$.  The potential is strongly enhanced for freestanding wires ($\epsilon_{e}=1$), whereas it is strongly damped for a high-$\kappa$ coating.}
\label{fig1}
\end{figure}

We consider an infinitely long semiconductor wire having a diameter of few nanometers. To calculate the electron mobility in such a structure, we first investigate the effect of dielectric mismatch on the ionized-impurity (donor) scattering rate. For a positive impurity ion situated on the axis of the wire (see Fig.\ref{fig1}), the Fourier transform of the bare electrostatic potential inside the nanowire can be written as \cite{smythe89, muljarovPRB00}
\begin{equation}
\tilde{V}^{Coul}(\rho,k) = \frac{e}{4\pi^{2} \epsilon_{s}}\Big[K_{0}(k \rho)+\frac{\pi\gamma}{2} e^{ - 2 k R } I_{0}( k\rho )\Big], 
\label{a}
\end{equation}
where $\gamma = (\epsilon_{s}-\epsilon_{e})/ (\epsilon_{s}+\epsilon_{e})$ is the dielectric mismatch factor, $I_{0}(...)$ and $K_{0}(...)$ are the zeroth order modified Bessel functions, $k = k_{z}$ is the electron wavevector along the wire axis, $\rho$ is the distance from the axis of the nanowire, $R$ is radius of the nanowire and $e$ is the electron charge. In Eq. \ref{a}, the second term arises due to the dielectric mismatch ($\gamma \neq 0$), and is a very good approximation of the exact potential for $| kR | > 1/4$.  
Momentum conservation rule allows only back scattering of carriers ($\vec{k} \rightarrow \vec{-k}$, $\vec{k}$ being the initial state) for transport in the first subband of the NW, and therefore results in a large momentum change ($q=2k$) in any elastic scattering process. So the assumption $| kR | > 1/4$ is justified.

The second term in equation \ref{a} captures the dielectric mismatch effect. The Coulomb potential experienced by a carrier electron
inside the wire is damped when $\epsilon_{e} \ge \epsilon_{s}$. For NWs with large radii, the transport properties in the nanowire should approach that of the bulk semiconductor; i.e., the dielectric mismatch should not affect the transport.  This fact is captured in Eq. \ref{a} - the mismatch term vanishes exponentially for large radii. In the long-wavelength limit, the real-space Coulomb potential experienced by an electron at $(\vec{\rho}, z), |\vec{\rho}|<R, z\gg R$ is found from Eq. \ref{a} to be $V^{Coul}(z) \approx  (e / 4 \pi \epsilon_{e} z )$, which indicates that Coulomb interaction within the wire is completely dominated by the {\em dielectric constant of the environment}, and the dielectric constant of the semiconductor does not have an effect on Coulomb interactions.

Assuming that the electrons are confined in an infinite barrier potential, the electron energies are $E_{n}(k)=E_{n}+\hbar^{2}k^{2}/2m^{\star}$, where $E_{n}$ is the ground-state energy of $n^{th}$ 1D subband, and $m^{\star}$ is the electron effective mass. The corresponding envelope function is $\Psi_{n,\vec{k}} (\vec{\rho},z)=\phi_{n}( \vec{\rho} ) \cdot [\exp(i k z) / \sqrt{L}]$, where $\phi_{n}(\vec{\rho})$ is the radial part, $\hbar k$ is the longitudinal quasi-momentum and $L$ is the length of the nanowire. We chose $E_{1}$ as the reference of energy. For thin nanowires the ground state radial part of the envelope function can be approximated by $\phi_{n}(\vec{\rho}) \approx 1/\sqrt{\pi R^2}$. Then the matrix element for Coulomb scattering from state $|0,k_{i}\rangle \rightarrow |0,k_{f}\rangle$ is calculated to be
\begin{equation}
\tilde{v}^{Coul}(q,R)=\frac{e^2}{ 2 \pi \epsilon_{s} L x^{2}} \Big[ 1-xK_{1}(x) + \frac{\pi\gamma x e^{-2x}}{2} I_{1}(x) \Big],
\label{b}
\end{equation}
where $x= qR = | \vec{k_{i}} - \vec{k_{f}} | R$.

In addition to Coulomb impurity scattering, electrons in III-V NWs also suffer from longitudinal-optical (LO) phonon scattering at room temperature. LO phonon scattering is adequately modeled by 3D modes even for nanowires \cite{fishmanPRB87}. Taking the phonon wave vector to be $\vec{q}_{ph}=(\vec{Q},q_{z})$, $\vec{Q} = (q_{x},q_{y})$, the electron-phonon scattering matrix element can be written as \cite{fishmanPRB87}
\begin{eqnarray}
\tilde{v}^{e-ph}(q,q_{z},Q,R)&=&\sqrt{\frac{ n_{ph}^{\pm} e^{2} \hbar \omega_{0}}{ 2 (Q^{2}+q_{z}^{2})}(\frac{1}{\epsilon_{s}^{\infty}} - \frac{1}{\epsilon_{s}^{0}})} \nonumber \\
&&\times\frac{2J_{1}(QR)}{QR}\delta_{q,\pm q_{z}},
\end{eqnarray}
where $\epsilon_{s}^{\infty}$ ($\epsilon_{s}^{0}$) is the high (zero) frequency dielectric constant of the semiconductor, $n_{ph}^{-} = [1+\exp{(\hbar \omega_{0} / kT)}]^{-1}$, $n_{ph}^{+} = (1 + n_{ph}^{-}) \Theta [ E_{n}(k) - \hbar \omega_{0} ]$ stand for absorption and emission of a LO phonon of energy $\hbar\omega_{0}$ respectively ($\Theta [...]$ is the Heaviside unit-step function), and $J_{1}(...)$ is the Bessel function. 

Using scattering matrix elements for the Coulomb and phonon scattering matrix defined in Eqs. 2 \& 3 , the scattering rate for the $i^{th}$
scattering mechanism  is calculated as 
\begin{eqnarray}
\frac{1}{\tau_{i}(k)}=
\frac{2\pi}{\hbar}\int\frac{dk'}{2\pi}|\frac{\tilde{v^{i}}}{\epsilon(q,0)}|^{2}(1-\cos\theta)\delta(E_{k}-E_{k'}\pm \hbar\omega_{0}),\nonumber\\
\end{eqnarray}
where $\pm\hbar\omega_{0}$ is required only for inelastic LO phonon scattering, and $\epsilon(q,0)$ is the screening factor in the static limit
($\omega\to 0$). For electron motion in the first subband only back scattering is possible, hence $\cos\theta = -1$. The scattering rate is summed up over the final density of states and the Drude mobility is given by $\mu = e\langle\tau(T)\rangle/m^{\star}$, where $\langle\tau(T)\rangle$ is the ensemble-averaged scattering rate
\begin{eqnarray}
\langle\tau(T)\rangle = \frac{\int_{0}^{\infty}dk k \tau_{m}(k)(-\frac{\partial f_{0}}{\partial k})}{\int_{0}^{\infty}dk f_{0}(k)}.
\end{eqnarray}
Here $f_{0}(k)$ is the Fermi-Dirac distribution, and $\tau_{m}(k)$ is the momentum relaxation time. 
\begin{figure}[t]
\includegraphics[width=5cm]{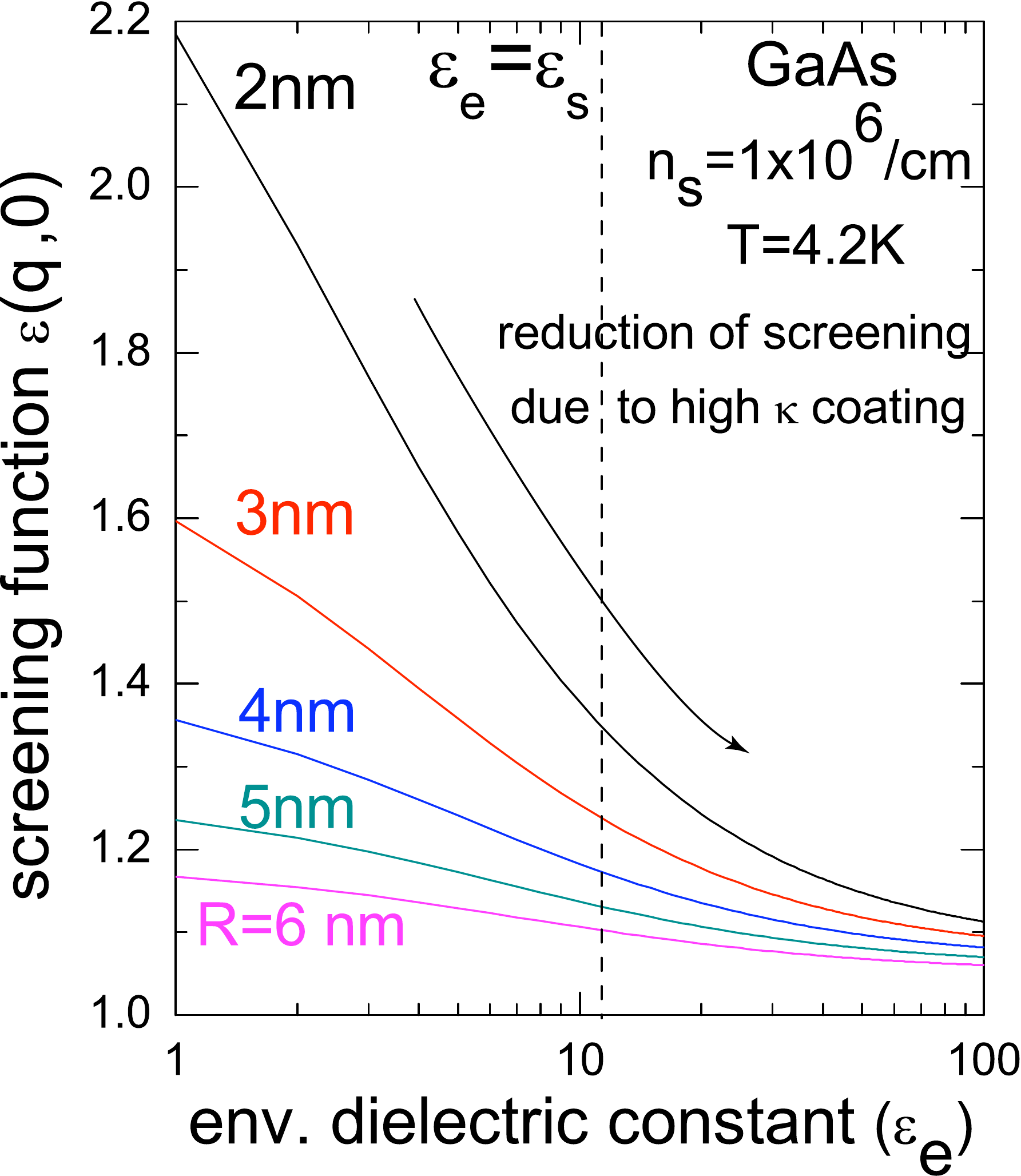}
\caption{\label{fig:epsart} Screening function for different nanowire radii as a function of environmental dielectric constant: for low dielectrics and small radii, screening is significantly strong.}
\label{fig2}
\end{figure}
The bare potential in NWs is screened due to the presence of free carriers. The quasi 1D screening function $\epsilon(q,0)$ can be calculated using the self-consistent procedure outlined by Lee and Spector \cite{leeJAP85}; we have included the effect of the dielectric mismatch to their screening theory. At low temperatures, the carriers contributing to transport are predominantly those at the Fermi level, and the momentum change upon scattering is $q \approx 2k_{F}$, where $k_{F}$ is the Fermi wavevector. It is well known that the 1D screening function diverges for $q = 2k_{F}$ at $T=0$ K. So we have used Maldague's \cite{maldagueSS78} prescription to remove the zero temperature singularity to obtain the finite temperature screening function as $\epsilon( q, 0) = 1 + \Pi (q, R, T)/q^{3}$, where
\begin{equation}
\Pi(q, R, T) = \frac{ \frac{1}{2} + I_{1}(v) \big[ \frac{ \pi \gamma e^{-2v} }{2} I_{1}(v) - K_{1}(v) \big]}{\pi R^{2} a_{B}^{\star}} S(u)
\end{equation}
where $v=2qR$, $u=\epsilon_{F}/k_{B}T$, and $a_{B}^{\star}$ is the effective Bohr-radius of the bulk semiconductor.  $S(u)$ is a dimensionless integral defined in \cite{fishmanPRB86}. Screening is strong for for a low-$\kappa$ dielectric coating when $qR\le1$.  The effect of free-carrier screening is found to be negligibly small for NWs coated with a high-$\kappa$ dielectric, as shown in Fig. \ref{fig2}.  This can be understood from electrostatics: the electric field lines prefer to bunch in regions of high dielectric constant to lower the energy.  Therefore, for a high-$\kappa$ coating around a thin NW, the field lines leak out into the surrounding and thus free-carrier screening becomes ineffective.  For large NW radius, as expected, the dielectric mismatch has a very weak effect on screening (see Fig. \ref{fig2}).

\begin{figure}[h]
\includegraphics[width=8cm]{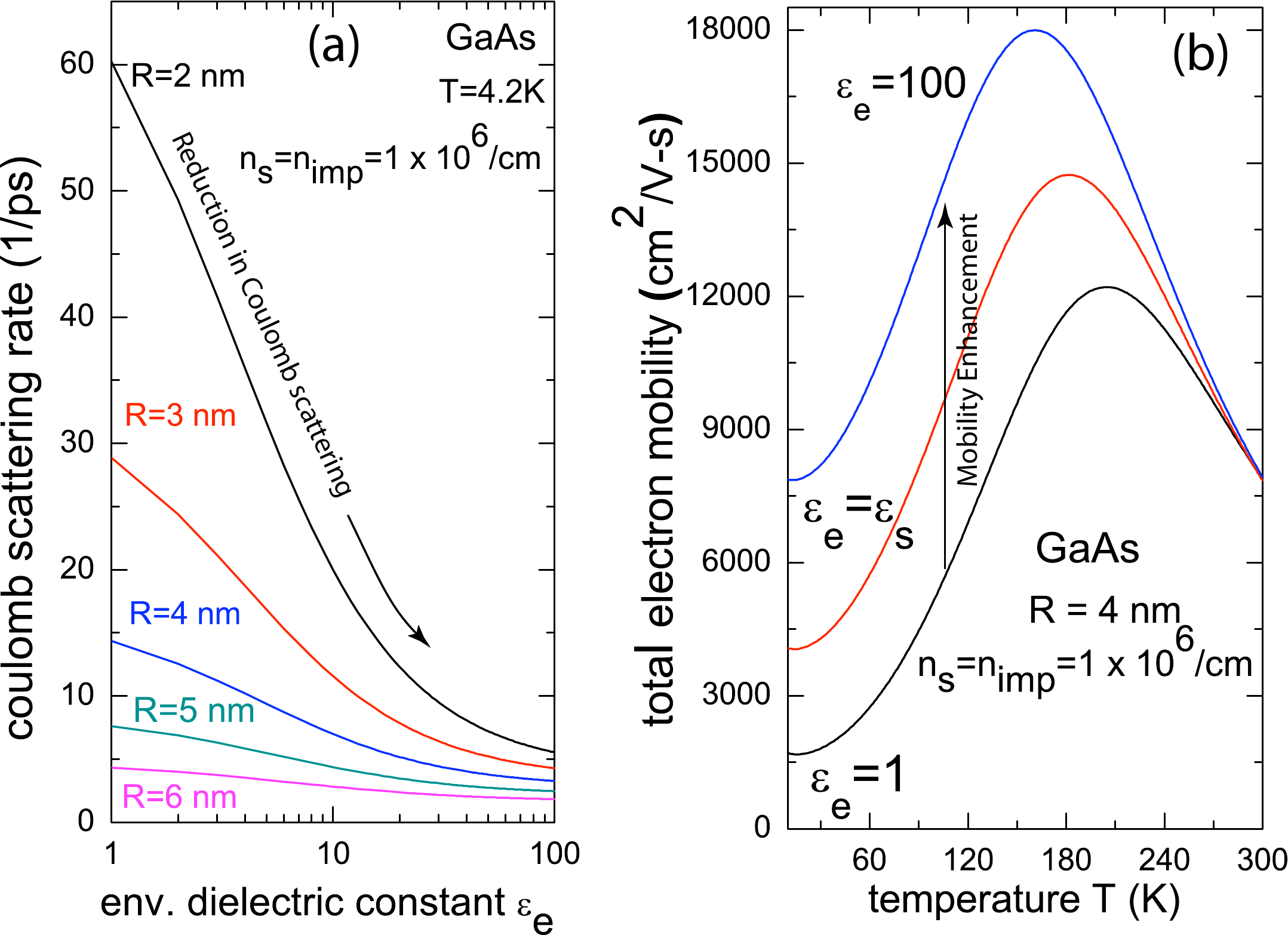}
\caption{\label{fig:epsart} a) Dependence of charged impurity scattering rate for various radii of NWs on the environmental dielectric constant. With increasing dielectric constant of the environment, the scattering rate decreases. b) electron mobility in a 4nm GaAs NW as a function of temperatures for different dielectric environments.} \label{fig3}
\end{figure}

For numerical evaluation of the effect of dielectric mismatch on transport, we assume a 1D electron density $n_{1D}=10^{6}$ cm$^{-1}$, and the ionized impurity density to be the same.  For this density, the Fermi energy for a doped GaAs NW is $E_{F} = 14$ meV.  Then, to ensure that transport occurs in the first subband, the radius should be $R< 20$ nm.  At low temperatures, $E_{F}\gg k_{B}T$ and $q \approx 2k_{F}$, where $k_{F}$ can be determined from the electron density ($ k_{F} = \pi n_{1D} / 2 $).  The calculated Coulomb scattering rates at $T=4.2$ K are shown in Figure \ref{fig3} (a) for a GaAs NW with radii varying from 2 - 6 nm.  For a 2 nm radius NW, the Coulomb scattering rate decreases from 60/ps to 6/ps when the dielectric constant if the environment is changed from 1 (air) to 100 (high $\kappa$ dielectric).  With increasing radius, Coulomb scattering rate becomes insensitive to the dielectric environment (see Eq.\ref{b}). 

For the calculation of temperature-dependent electron mobility, phonon scattering must be considered. Typical values for GaAs have been used ($\hbar\omega_{0}=36 $ meV, $\epsilon_{s}^{\infty} = 11 \epsilon_{0}$, and $\epsilon_{s}^{0} = 13 \epsilon_{0}$). Phonon scattering rates are weakly dependent on the dielectric mismatch ($\gamma$) through the screening function.  At low temperatures optical phonon scattering is negligible compared to Coulomb scattering rate for our choice of impurity density. So the carrier mobility in the nanowire is determined by Coulomb impurity scattering. Therefore, the electron mobility is strongly dependent on the dielectric environment, as shown in Figure \ref{fig3}(b); for example, coating the GaAs NW with a high $\kappa$ dielectric can result in as much as 4 times enhancement in carrier mobility as compared to a freestanding nanowire due to the large damping of Coulomb scattering.  At higher temperatures the LO phonon scattering rate increases and at room temperature (T=300 K) LO phonon scattering dominates over Coulomb scattering even for high impurity densities. As a result dielectric mismatch effect on the carrier mobility vanishes at room temperature.  We point out here that in NWs made of elemental semiconductors such as Si or Ge, LO phonon scattering is weak, and for heavy doping, the dielectric mismatch induced enhancement in carrier mobility should persist up to room temperature.  In comparison to semiconductor nanomembranes, the Coulomb scattering rate in wires is damped due to the reduced density of states near the Fermi energy, and therefore the drastic dielectric effect expected for nanomembranes \cite{jenaPRL07} is not observed for doped III-V nanowires.
Two questions that have not been addressed in this work are a) the effect of charged surface states on nanowires, and b) the effect of surface roughness scattering.  These are equally important questions since the presence of a large density of surface states on the nanowire will alter the electrostatic boundary conditions and hence lead to different Coulomb potential.  Similarly, surface roughness scattering can compete with Coulomb scattering at low temperatures, and possibly even dominate for very thin wires with rough surface morphologies.  These extensions to the core model presented here will be presented in a more comprehensive later work.

In conclusion, we have investigated the effect of the dielectric environment on the electron mobility in semiconductor NWs. It is found that the Coulomb potential inside the nanowires can be tuned the dielectric environment.  Coating a thin NW with a high-$\kappa$ dielectric will damp the Coulomb scattering, and if charged impurity scattering is the dominant scattering mechanism, mobility can indeed be improved by a high-$\kappa$ coating around the NW.  This is a novel technique for enhancing the mobility in nanostructures, and is well suited for applications in Field-Effect Transistor structures, where a high-$\kappa$ dielectric affords better gate control, in addition to a higher mobility as shown in this work.

\bibliography{exact_wire_last_v1}

\end{document}